\documentclass[nofootinbib,prd,superscriptaddress,article,showkeys]{revtex4}
\usepackage{amsmath,amssymb,graphics,epsfig}
\usepackage{enumerate}
\usepackage{theorem}

\begin{document}


\title{\huge Relativistic Finsler geometry}




\author{A.P.Kouretsis }
\email{ akouretsis@astro.auth.gr}
\thanks{Corresponding author}
\affiliation{Section of Astrophysics, Astronomy and Mechanics, Department of Physics Aristotle University of Thessaloniki, Thessaloniki 54124, Greece}

\author{M.Stathakopoulos}
\email{michaelshw@yahoo.co.uk}
\noaffiliation

\author{P.C.Stavrinos}
\email{pstavrin@math.uoa.gr}
\affiliation{Department of Mathematics, University of Athens, Athens 15784, Greece}

\begin{abstract}
We briefly review some  basic concepts of parallel displacement in Finsler geometry. In general relativity,  the parallel translation of objects along the congruence of the fundamental observer  corresponds to the evolution in time. By dropping the quadratic restriction on the measurement of an infinitesimal distance, the geometry is generalized to a Finsler structure. Apart from curvature a new property of the manifold complicates the geometrodynamics, the color.   The color brings forth an intrinsic local anisotropy and many quantities depend on position and to a "supporting" direction. We discuss  this direction dependence and some physical interpretations. Also, we highlight that in Finsler geometry the parallel displacement isn't necessarily always along the "supporting" direction. The latter is a fundamental congruence of the manifold and induces a natural $1+3$ decomposition. Its internal deformation is given through the  evolution of the irreducible components of vorticity, shear and expansion.
\end{abstract}



\maketitle

During the onset of general relativity  possible geometric extensions have been considered mainly by introducing extra dimensions and/or torsion to space-time. In our days such theories are most likely interpreted as effective theories of stringy models and are included in the field of quantum gravity phenomenology. Many scenarios of string theory lead to vacuum expectation values with non-zero tensor fields. Consequently, the local structure of space-time brakes down from the view point of general relativity since the 4D Minkowski vacuum does not accurately describe any region of the "physical" manifold. The non-zero tensorial entities are expected to violate the local spatial isotropy and will introduce space-time asymmetries even in asymptotic regions of space-time. This  reminds us an anisotropic medium in place of the standard flat solution and is closely connected to the "analogue spacetime" programme \cite{Barcelo:2005fc}. From a geometric standpoint one may generalize his metric structure to include such effects \cite{Miron}. The local anisotropy may effectively interpreted as a stochastic quantity  or through a collective behavior as a long range tensorial entity. Riemann geometry can play a crucial roll since it may stand as a background metric of an asymptotic state while the "physical" geometry will be interpreted by a foreground anisotropic metric.

Breaking the local symmetries of space-time creates an incompatibility with the 4D Riemann space-time \cite{Kostelecky:2003fs} and the tangent space will be characterized by intrinsic variables. This is clearly monitored in the bimetric structures where the disformal correlation between the foreground metric and the Riemann structure induce a distorted unit ball at each tangent space ({\it indicatrix}), even if we replace the background metric with a Minkowski space-time. A candidate geometry to describe such structures is Finsler geometry since the properties of the base manifold depend on position and direction; or as it is usually called   "{\it element de support}" \cite{RundFB}. However, the direction dependence falls under certain restrictions as for example it must not be a null vector field \cite{Skakala:2010hw}. The anisotropic unit direction (for now on we will call it {\it supporting element}) is considered to be an autoparallel bundle and through standard variational principles we can prove that it coincides with the extremal curves for almost any Finslerian connection.

The generalization to Finsler manifolds is achieved by simply dropping one of the fundamental conditions of Riemann geometry. In particular, we drop the restriction that the line element
\begin{equation}
ds=F(x,dx)\label{ds},
\end{equation}
should depend only on quadratic terms with respect to the infinitesimal coordinate increments, $dx^a$. Yet the metric function $F(x,dx)$ falls under all the other restrictions of Riemann geometry. The most significant among them, is the statement that the distance between two points remains invariant under reparametrization of the connecting curve, namely
  \begin{equation}
  \int F(x,\frac{dx}{dt})dt=\int F(x,\frac{dx}{ds})ds.\label{Fmet}
  \end{equation}
   This property is achieved by demanding that the metric function is homogeneous of first degree with respect to the coordinate increments, $F(x,\lambda dx)=\lambda F(x,dx)$. Taking advantage of the latter we introduce the Finslerian metric tensor
  \begin{equation}
  g_{ab}(x,y)=\frac{1}{2}\frac{\partial^2F^2}{\partial y^a \partial y^b},
  \end{equation}
  that reduces to the Riemannian metric when relation (\ref{ds}) is the quadratic form $ds^2=a_{ab}(x)dx^adx^b$. In the most general case the resulting geometry is a fiber space of $(x,y)$ where $y^a$ is an arbitrary direction. This point of view corresponds to Cartan's approach to the subject and brings forth a metric geometry with torsion \cite{Ish}. However, when we set $y^a$ to be the fundamental congruence (observer) along which we evolve various entities, the geometric formulas are suitably simplified.
 The adoption of a non-quadratic measurement enrich the manifold with an extra property, the color \cite{Shen}. It monitors the deviation from the quadratic measurement and it may vary from region to region. In that case we say that the manifold is colorful. In simpler cases, the color may be constant \footnote{When the manifold is mono-colored we get the Berwald geometry that still satisfies the axiomatic approach to gravity of Ehlers, Pirani and Schild \cite{Tavakol}.} and when the geometry is Riemann the manifold becomes {\it white}.   The structure equations (Bianchi identities) of Finsler geometry reflect the colorful "morphology" and give us information for the color, the curvature as well as the interplay between them. The supporting element emerges from the color of Finsler and in the general case plays the role of an internal variable. However, we can also study the properties of the supporting element itself as a geodesic congruence.

The key question in Finsler geometry comes from the notion of parallel displacement \cite{Hach}. In the general case we can consider the parallel displacement of a vector field along a geodesic with respect to the {\it supporting element}. In that case, the propagation along the congruence involves notions from the geometry of vector bundles. Precisely, the dragging along the geodesic operates in the tangent bundle of the total space, $\mathcal{TTM}$, as long as the supporting element is considered undetermined with respect to the position coordinates. However, we can imagine restrictive cases for the supporting element to retrieve less complicated structures. In fact, for almost any connection definition the parallel displacement {\it along the element of support} is given by an operator that is a direct generalization of the usual covariant derivative.

The "{\it element de support}" defines a geodesic congruence with analogous properties to the Riemannian geodesics. Also, for a $n$-dimensional manifold it introduces a natural $1+(n-1)$ split since we can decompose tensor fields to their irreducible parts along  and on the transverse (n-1)-hypersurface of the supporting direction \cite{Shen, Matsum}. In particular the {\it elemental} decomposition is done with the aid of the following tools
\begin{equation}
l_a=\frac{\partial F}{\partial y^a}\;,\;\;\;h_{ab}=F\frac{\partial^2F}{\partial y^a\partial y^b},\label{1k3}
\end{equation}
where $l_a$ is the {\it normalized supporting direction} while $h_{ab}$ projects orthogonal to $l_a$ and is called the {\it angular metric tensor}. Precisely, recalling the first order homogeneity of $F(x,y)$ we get from the previous relation  $l_a=y^a/F$. For the same reason we can prove that $h_{ab}y^b=0$ and that the rank of $(h_{ab})$ is $n-1$. Finally, we can recast the angular tensor in the convenient form
\begin{equation}
h_{ab}=g_{ab}-l_al_b,
\end{equation}
where $l^al_a=1$ is a unit time-like vector. Apparently, we may decompose tensor fields to their irreducible parts with respect to the {\it normalized supporting element} by the following process
\begin{equation}
X_a=g_a^{\;\;b}X_b=(h_a^{\;\;b}+l_al^b)X_b=Xl_a+\mathcal{X}_a
\end{equation}
where $X=X_al^a$ is the longitudinal part while $\mathcal{X}_a=h_a^{\;\;b}X_b$ is the projected component orthogonal to $l_a$.
Concerning physical applications of Finsler geometry in relativity, the role of the supporting element needs to be clearly defined. It may stand for a local anisotropic variable that may reflect a Lorentz symmetry breaking (or generally speaking a "wind" that exists in the foreground space-time), or  the velocity of the fundamental observer in relativistic kinematics. The latter case is  closely related to the Riemannian case since as we will see in the following the parallel translation along the {\it supporting element} $l^a$ (in other words the evolution) shares many similarities with the Riemannian case \cite{RundFB}.  In the literature one can find all this plausible scenarios where Finsler geometrodynamics stands as an effective description of the foreground manifold or as a fundamental theory of extended gravitational physics.

{\it In the following we overview some concepts of parallel translation in Finsler geometry using the Cartan Euclidean connection. Then we restrict the supporting element to be the velocity of the fundamental observer in analogy to the $1+3$ covariant formalism of relativistic cosmology.}

As we mentioned in the discussion, in Finsler geometry tensor fields depend on position and direction. Technically speaking for a first rank tensor we can write down
\begin{equation}
X_{a}\equiv X_{a}(x,y),
\end{equation}
where $(x,y)$ is the "{\it element de support}", $y^a$ stands for the tangent vector ${dx^a}\over d\tau$ of an extremal bundle $\sigma(\tau)$ and $x^a$ is the coordinate net of the base manifold. The usual coordinate transformations are valid for any tensor field, $\tilde{X}^a={\partial \tilde{x}^a\over \partial x^b}X^b$.  In the footsteps of Euclidean geometry the parallel displacement of $X^a(x,y)$ along an arbitrary congruence $\gamma(s)$ is
\begin{equation}
D_{(\gamma)}X^a=dX^a+\Gamma^a_{\;\;bc}(x,y)X^bdx^c+C^a_{\;\;bc}(x,y)X^bdy^c\label{Dx},
\end{equation}
where the index in the $D_{(\gamma)}$ operator denotes that we evolve $X^a$ along $\gamma(s)$. Note, that in the general case $\sigma(\tau)\neq\gamma(s)$. Moreover, additional geometric postulates indicate explicit formulas for the coefficients $\Gamma^a_{\;\;bc}$ and $C^a_{\;\;bc}$. The last term of the previous expression reflects the internal anisotropy of the Finslerian manifold and brings into play the structure of $\mathcal{TTM}$. The dependence on the "{\it element de support}" is a direct consequence of our departure from the quadratic measurement of the $y^a$-norm, $F(x,y)$.

One of the most dominant perspectives is  Cartan's Euclidean connection that implies a metric theory. In this approach the length of a parallel displaced vector along any direction remains constant. We can recast relation (\ref{Dx}) in a more convenient way using the following covariant differential operators
\begin{equation}
\frac{DX^a}{ds}\equiv X^a_{\;\;|b}\frac{dx^b}{ds}+X^a|_{b}F\frac{Dl^b}{ds},\label{covX}
\end{equation}
where the small bar ${}_|$ is  the covariant derivative of the horizontal subspace ($h$-derivative), the large bar $|$ stands for the covariant derivative of the vertical subspace ($v$-derivative) and $l^a=y^a/F$ is the {\it unit supporting element} (for further details see \cite{Miron,RundFB}). The relation (\ref{covX}) monitors the parallel displacement of a first rank tensor along the $\frac{dx^a}{ds}$-congruence with respect to the {\it unit supporting direction}.

 The two different covariant derivatives result from the dependence of tensorial entities on tangent bundle variables. Therefore, the differentiation takes place in the vector bundle $\mathcal{TTM}$. They reflect the natural split of $\mathcal{TTM}$ to the horizontal and vertical subbundles given by the Whitney sum
\begin{equation}
\mathcal{TTM}=\mathcal{HTM}\oplus\mathcal{VTM}.
\end{equation}
The above split comes from the nonlinear connection distribution
$H:u\in\mathcal{TM}\rightarrow \mathcal{H}_{u}\mathcal{TM}\subset\mathcal{T}_u\mathcal{TM}$
which is equivalent to the local expression $N_b^{\;\;a}=\frac{\partial G^a}{\partial y^b}$. The coefficients $G^a$ are the {\it spray} defined by the extremal curves of the {\it supporting element}
\begin{equation}
\frac{Dy^a}{d\tau}=\frac{dy^a}{d\tau}+2G^a(x,y)=0,\label{spray}
\end{equation}
where (\ref{spray}) is an identical expression to the Riemannian geodesics for the usual Christoffel symbols  $\gamma^a_{\;\;bc}$ with  only difference the dependence of the metric in $(x,y)$. We can also reexpress the spray coefficients with respect to the norm $F(x,y)$, as
\begin{equation}
G^a={1\over4}g^{ab}\left(\frac{\partial^2F^2}{\partial x^c\partial y^b}y^c-\frac{\partial{F^2}}{\partial x^b}\right)\label{spco}
\end{equation}

Concerning the second variation of an arbitrary integral curve, the covariant expression (\ref{covX}) suggests that there will be three different curvature tensors generated by the mixture of the $h$ and $v$-derivatives \cite{Miron,RundFB}. However, by virtue of relation (\ref{spray}), when we restrict the displacement of $X^a$ along the {\it supporting element}  the vertical part of (\ref{covX}) is switched off restraining the variation on the horizontal subbundle. In that case the parallel displacement of $X^a$ along the supporting direction is given by the following formula,
\begin{equation}
\frac{DX^a}{d\tau}\equiv X^a_{\;\;|b}\frac{dx^b}{d\tau}=X^a_{\;\;|b}y^b.\label{hcovX}
\end{equation}
The $h$-derivative brings forth the effect of color on the evolution of $X^a$ along $y^a$ through the non-linear connection. Furthermore, if $X^a$  is the supporting element , $X^a= y^a$, the absolute differentiation (\ref{Dx}) along $\sigma(\tau)$ reduces to the expression (\ref{spray}) for the extremal curves,
 \begin{equation}
 \dot{y}^a\equiv\frac{Dy^a}{d\tau}=y^b\nabla_by^a\label{auto}
 \end{equation}
 where the covariant derivative $\nabla_a$ is an analogous expression to the Levi-Civita connection \footnote{When we contract various Finsler quantities along $y^a$ they reduce to simpler forms. In particular, the autoparallel curves take the same expression as in Riemann geometry for the Cartan Euclidean connection as well as for the Chern-Rund connection, $\Gamma^a_{\;\;bc}y^by^c=\gamma^a_{\;\;bc}y^by^c$. Nevertheless, the dependence of the metric on the "{\it element de support}" still remains.}. From now on the dot operator on a tensor field  ${\dot{\mathcal{Z}}^a}$,  denotes the evolution (parallel displacement) along a given congruence. Note, that the autoparallel curves (\ref{auto}) coalesce with the extremal curves\footnote{The same argument holds for the Chern-Rund connection.} (\ref{spray}).

Noteworthy is the limiting case $y^a=y^a(x)$, known as the osculating Riemannian limit \cite{RundFB,Asanov,Oscul}. This restricts our analysis to a subregion of the Finsler manifold $F_n$ where the metric tensor and any geometric entity depends solely on the position,
 \begin{equation}
 r_{ab}(x)=g_{ab}(x,y(x)).
 \end{equation}
 However, from a physical viewpoint $y^a$ can play again either the role of the fundamental observer in relativistic kinematics or the role of an internal variable. In the first case the standard Levi-Civita differentiation holds and the geodesics coincide with the geodesics of a Riemannian manifold while in the second case a non-Riemann space with torsion is introduced. Precisely, if we impose that the absolute differentiation is always along the {\it supporting element} $y^a$ then we retrieve the standard relation for the geodesics,
 \begin{equation}
 \frac{dy^a}{d\tau}+\frac{1}{2}r^{ab}(r_{bc,d}+r_{bd,c}-r_{cd,b})y^cy^d=0,
 \end{equation}
 and for the osculating subregion they coincide with the extremal curves of a Riemannian manifold. This  implies the limit of relativistic cosmology and gives useful insight for the imprint of color on the cosmological fluid \cite{Oscul}. For particular metric functions the cosmo-dynamics will be modified providing a way to "loosely" constrain the Finslerian parameters.

 On the other hand, when $y^a$ corresponds to an internal variable another osculating structure can be retrieved by the non-integrable in general condition $dy^a=Z^a_{\;\;b}dx^b$. Then we can recast relation (\ref{Dx}) in the following formula\cite{RundFB,Ikeda}
 \begin{equation}
 D_{(\gamma)}X^a=X^a_{\;\;||b}dx^b
 \end{equation}
where the double bar is a covariant derivative with torsion
\begin{equation}
X^a_{||b}=\frac{{\overline{\partial}X^a}}{{\partial x^b}}+\overline{\Gamma}^a_{\;\;bc}X^c
\end{equation}
 and we define $\overline{\Gamma}^a_{\;\;bc}=\Gamma^{a}_{\;\;bc}+Z^d_{\;\;c}C^{a}_{\;\;bd}$ with $\frac{\overline{\partial}}{\partial x^a}=\frac{\partial}{\partial x^a}+Z^b_{\;\;a}\frac{\partial}{\partial y^b}$. The covariant formulas of this structure are the same with the formulas of Einstein-Cartan geometry. For example the Ricci identities read
\begin{equation}
X^a_{\;\;||b||c}-X^a_{\;\;||b||c}=\overline{R}^a_{dbc}X^d-\overline{Z}^d_{\;\;dc}X^a_{\;\;||d}\label{Rtor},
\end{equation}
with $\overline{R}_{abcd}$ standing for the curvature tensor given by the usual formula for the overline definitions. The last term of relation (\ref{Rtor}) involves a covariant 3rd rank tensor resulting from the antisymmetry of the connection
\begin{equation}
\overline{Z}^a_{\;\;bc}=\overline{\Gamma}^a_{\;\;bc}-\overline{\Gamma}^a_{\;\;cb},
\end{equation}
with the same properties to the classic torsion tensor. The torsion field $\overline{Z}^a_{\;\;bc}$ is a direct effect of the non-quadraticity of the metric function $F(x,y)$. In other words in this limit the color of Finsler induces torsion to space-time. From a physical perspective this introduces a curvature theory compatible with Lorentz violations. In this framework, we may consider a stochastic distribution for $y^a$ as a space time fluctuation that will introduce torsion through relation (\ref{Rtor}) to the kinematics of a relativistic fluid (see the related article \cite{Mavromatos:2010ar}). All this options of parallel translations open up a wide arena of geometrodynamics that can find place inside classic general relativity as well as  inside extended theories of gravity.


 Now let us concentrate our analysis to the internal deformations of the geodesic bundle defined by the {\it supporting element}.  Consider a nearby geodesic to the supporting element's integral curves and the connecting vector $\xi^a$ given by
\begin{equation}
\tilde{x}^a=x^a+\epsilon\xi^a,\label{inftr}
\end{equation}
where $\epsilon$ stands for a small parameter. Then, by virtue of relation (\ref{inftr}) and for a first order approximation with respect to $\epsilon$, the Euler-Lagrange equations (\ref{spray}) imply the following expression for the deviation of geodesics
\begin{equation}
\ddot{\xi}^a+\mathcal{H}^a_{\;\;b}(x,y)\xi^b=0.\label{dev}
\end{equation}
The above expression tracks the relative displacement between two neighboring integral curves of the {\it supporting element} bundle. The second rank tensor $\mathcal{H}^a_{\;\;b}$ is the {\it relative curvature} with respect to the supporting direction. In Finsler geometry is non-trivially correlated to the tangent of the supporting geodesic congruence. We remind the reader that the previous relation is not the most general case  since the translation is along $y^a$. Hence, many properties of Riemann geometry still hold.

The eigenvalues of $\mathcal{H}^a_{\;\;b}$ are the principal sectional curvatures that characterize the deviation between two neighboring geodesics \footnote{We remark that $\mathcal{H}^a_{\;\;b}$ is of the same form either in case of Cartan or Chern-Rund connection \cite{RundFB}. In other words the Jacobi equation is of the same form in both cases. For some recent discussion on the Finslerian generalization of the Jacobi equation using the Chern-Rund connection see  \cite{Torrome}.}. Its explicit form is given with respect to the spray coefficients (\ref{spco}) and the non-linear connection, namely
\begin{equation}
\mathcal{H}^{a}_{\;\;b}\equiv2\frac{\partial G^a}{\partial x^b}-y^c\frac{\partial N^a_{\;\;b}}{\partial x^c}+2G^c\frac{\partial N^a_{\;\;c}}{\partial y^b}-N^a_{\;\;c}N^c_{\;\;b}\label{relH}.
\end{equation}
The deviation equation (\ref{dev}) is related to the Riemannian curvature tensor derived from the horizontal covariant derivative
\begin{equation}
\mathcal{H}_{ab}=R_{acbd}(x,y)y^cy^d,
\end{equation}
where in the Riemannian limit $R_{abcd}$ depends solely on the position coordinates. From a relativistic standpoint, relation (\ref{dev}) describes the relative acceleration between neighboring observers. The acceleration is in the foreground manifold where the gravitational force may not be the only tidal effect \cite{Barcelo:2005fc}.

To demonstrate this let as consider the Riemannian metric $\alpha=\sqrt{\alpha_{ab}y^ay^b}$ together with a 1-form $\beta=b_ay^b$ \cite{Matsum}. Assuming a Randers metric function
$F(x,y)=\alpha+\beta$, relation (\ref{Fmet}) gives back the explicit expression for the fundamental tensor
\begin{equation}
g_{ab}={F\over\alpha}(\alpha_{ab}-V_aV_b)+l_al_b,\label{Rmetr}
\end{equation}
where we define  the normalized vector with respect to the Riemannian metric $V_a={1\over\alpha}\alpha_{ab}y^b$. In Randers spaces we can extract formulas related to the Riemannian subregion. This is clearly depicted in the previous relation where a disformal correlation between the foreground metric $g_{ab}$ and the background $\alpha_{ab}$ is given. The Euler-Lagrange equations imply the Finslerian geodesics
\begin{equation}
y^a_{\;\;;b}y^b+2F^a_{\;\;b}y^b=0.\label{mf}
\end{equation}
 We denote by $;$ the usual Riemannian covariant derivative of the $\alpha$ metric, and $F_{ab}=b_{[a,b]}$ is a purely antisymmetric tensor. The last term in the lhs of the above relation interprets the Finsler geodesics as a magnetic flow of a Riemannian space and brings forth an interesting interplay between geometry and physics. For example, in three dimensions, a Randers structure of constant flag curvature corresponds to a Killing magnetic field \cite{GibZerm}.
 \footnote{The bimetric relation (\ref{Rmetr}) provides a natural framework to study motion in anisotropic media. In fact, for $l_a=\ell V_a$ it is conformally related to the optical metric of a medium with a varying optical index
 \begin{equation}
 g_{ab}={F\over\alpha}(\alpha_{ab}+{\beta\over\alpha}V_aV_b)\label{Rsmetr}.
 \end{equation}
In that case the 1-form $\beta$ that brakes the quadratic measurement is directly connected to the refractive index $n$ of the medium $1-n^{-2}={\beta\over\alpha}$. We remark also that relation (\ref{Rsmetr}) implies that the generalized Gordon ansatz in massive and bimetric gravity may be interpreted as a Finsler space-time\cite{VisBimetric}. Also, a similar conformal metric  appears as the foreground metric of perturbations on potential flows in arbitrary gravitational fields \cite{acretion}, for
\begin{equation}
{\beta\over\alpha}=1-u_s^2=1-\left({\partial P\over \partial\rho}\right)_s,
\end{equation}
where a Finsler structure naturally emerges once again. At this point, we remind the reader that the Finsler metric tensor is homogeneous of degree zero with respect to the coordinate increments. In fact, in all the above cases we deal with normalized vector fields that guarantee the 0-homogeneity of $g_{ab}$. }

The deviation equation (\ref{dev}) monitors the internal deformation of the fundamental congruence  $l^a$. The curvature tensor $\mathcal{H}_{ab}$ is symmetric and purely spatial, $\mathcal{H}_{ab}l^b=0$. Furthermore, using the covariant expressions (\ref{1k3}) we can decompose the relative curvature to its irreducible parts,
\begin{equation}
\mathcal{H}_{ab}={1\over3}\mathcal{K}h_{ab}+\mathcal{H}_{\langle ab\rangle}.
\end{equation}
The first term is the scalar flag curvature of Finsler geometry and corresponds to the usual formula, $\mathcal{K}=R_{ab}(x)l^al^b$, when the manifold is white. The second term is the symmetric, projective and trace free part, $\mathcal{H}_{\langle ab\rangle}=\mathcal{H}_{(ab)}-{1\over3}\mathcal{K}h_{ab}$. It involves in the kinematics the Weyl curvature which in Finsler geometry is suitably defined but is a more complicated formula.
We can isolate particular properties of the normalized elemental bundle by introducing the distortion tensor,
\begin{equation}
\dot{\xi^a}=B^a_{\;\;b}\xi^b\label{dis},
\end{equation}
 and along the supporting direction is trivial to prove that $B_{ab}=\nabla_{b}l_a$. Then, by taking the time derivative of (\ref{dis}) and substituting in (\ref{dev}) we recover the evolution of the tensorial distortion
 \begin{equation}
 \dot{B}_{ab}+B_{ac}B^c_{\;\;b}=-\mathcal{H}_{ab}.\label{edis}
 \end{equation}
Within the Riemannian context,  relation (\ref{edis}) plays a keynote role in studies of cosmological fluids and gravitational collapse. It is also crucial in the formulation of various singularity theorems.

We can further decompose the distortion tensor relative to the fundamental observer $l^a$, as follows
\begin{equation}
\nabla_bl_a=\frac{1}{3}\theta h_{ab}+\sigma_{ab}+\epsilon_{abc}\omega^c,
\end{equation}
where we define with respect to the projected covariant derivative ${\rm D_a=h_a^{\;\;b}\nabla_b}$ the expansion scalar $\Theta={\rm D}^al_a$ that monitors volume changes , the shear $\sigma_{ab}={\rm D}_{\langle a}l_{b\rangle}$ that tracks shape distortions and  for the 3D Levi-Civita symbol, $\epsilon_{abc}=\epsilon_{abcd}l^d$, the  vorticity $\omega_a=\epsilon_{abc}{\rm D}^{b}l^c/2$ that corresponds  to the change of orientation \cite{Grraych}. The vorticity, shear and expansion that we defined are for  an infinitesimal volume element parallel transported  along the supporting direction $l^a$. Using the metricity condition $\dot g_{ab}=0$ we can further decompose relation (\ref{edis}) into its irreducible parts. The scalar part corresponds to the Raychaudhuri equation
\begin{equation}
\dot{\Theta}+{\frac{1}{3}}\Theta^2=-\mathcal{K}-2(\sigma^2-\omega^2),\label{ray}
\end{equation}
the projective, symmetric and trace free part gives back the evolution of shear
\begin{equation}
\dot{\sigma}_{\langle ab\rangle}=-\frac{2}{3}\Theta\sigma_{ab}-\sigma_{c\langle a}\sigma^c_{\;\;b\rangle}-\omega_{\langle a}\omega_{b\rangle}-\mathcal{H}_{\langle ab\rangle},
\end{equation}
closely related to the propagation of tensor perturbations and to anisotropies; and finally the propagation of vorticity
\begin{equation}
\dot{\omega}_a=-{2\over3}\Theta\omega_a+\sigma_{ab}\omega^b.
\end{equation}
The above relations have a direct analogy with the Riemannian case. Hence, the same arguments will hold for the kinematics of the $l^a$-congruence. For example, curvature is not a source of vorticity and irrotational geodesics with positive scalar flag curvature $\mathcal{K}>0$ always forms a caustic. Note, that this brings forth a beautiful connection between rigidity and singularity theorems of Finsler geometry \footnote{For a discussion of the deformable kinematics of the tangent bundle using the horizontal and vertical split of $\mathcal{TTM}$ see \cite{Stavrin}. Also, see \cite{Kouretsis:2012ys} for a $1+3$ covariant treatment of Finsler flows in an arbitrary direction with respect to the supporting element.}.

 Consider an $(\alpha, \beta)$ metric where $\alpha$ stands for a Riemannian metric in the background while $\beta$ is a 1-form. The flag curvature will depend on the Riemann curvature tensor of $\alpha$ and on $\beta$. For example, in a Randers space there is a simple correlation
\begin{equation}
\mathcal{H}_{ab}=f_{ab}(\mathcal{R},\beta)\label{Rh},
\end{equation}
where $f_{ab}$ is a tensor that depends on the full Riemann tensor of the background, $\mathcal{R}_{abcd}(x)$, and  can be explicitly calculated \cite{Shen}. Therefore, as we expected the background curvature non-trivially sources deformations in the fundamental congruence of the foreground space-time. For example we can prove that the Weyl curvature of the background will affect the expansion of the flow \footnote{This brings in mind the ideal MHD limit where the Riemannian curvature is non-trivially involved in the expansion dynamics through a magnetocurvature coupling. There, the Weyl curvature also enters the Raychaudhuri's equation \cite{Tsagas}. Similarities are expected since from relation (\ref{mf}) Finsler-Randers geodesics can be  physically interpreted as magnetic flows on a Riemannian manifold.}. The Finslerian kinematics of the medium may be visualized as the effect of classic Riemannian curvature on the internal deformation  of a congruence moving in a wind. The  $(\alpha, \beta)$ metrics induce an $f(R)$-like theory of gravity where we modify the way that the curvature of the background metric $\alpha$ generates deformations in a geodesic congruence of the foreground space-time (see relation (\ref{Rh}) combined with (\ref{edis})).  Of great interest is the effect on the expansion of the flow since Raychaudhuri's equation is directly related to the cosmological evolution. Precisely, positive terms in the rhs of relation (\ref{ray}) accelerate the expansion. {\it Thus, in case of an irrotational and shear-free "supporting" fundamental congruence, negative scalar flag curvature, $\mathcal{K}<0$, implies an accelerating expansion}.

\end{document}